\documentclass[twocolumn,preprint]{rmaa}

\usepackage{epstopdf}
\usepackage{textcomp}
\usepackage{multirow} 

\title{Radio and Optical Observations of DG Tau B}

 \author{
Luis F. Rodr\'\i guez\altaffilmark{1,2},
Sergio A. Dzib\altaffilmark{1},
Laurent Loinard\altaffilmark{1,3},
Luis A. Zapata\altaffilmark{1},
Alejandro C. Raga\altaffilmark{4},
Jorge Cant\'o\altaffilmark{5}
and
Angels Riera\altaffilmark{6}}
  \altaffiltext{1}{Centro de Radioastronom\'{\i}a y Astrof\'{\i}sica, UNAM, Campus Morelia}

\altaffiltext{2}{Astronomy Department, Faculty of Science, King Abdulaziz University, P.O.
Box 80203, Jeddah 21589, Saudi Arabia}

\altaffiltext{3}{Max-Planck-Institut f\"ur Radioastronomie, Bonn}

\altaffiltext{4}{Instituto de Ciencias Nucleares, UNAM}

\altaffiltext{5}{Instituto de Astronom\'\i a, UNAM}

\altaffiltext{6}{Departament de F\'\i sica i Enginyeria Nuclear, EUETIB, Universitat Polit\`ecnica de Catalunya}

\fulladdresses{
\item Sergio A. Dzib, Luis F. Rodr\'\i guez, Luis A. Zapata: Centro de Radiostronom\'{\i}a
y Astrof\'\i sica, 
UNAM, A. P. 3-72, 
(Xangari), 58089 Morelia, Michoac\'an, M\'exico
(s.dzib, l.rodriguez, l.zapata@crya.unam.mx).
\item Laurent Loinard: Max-Planck-Institut f\"ur Radioastronomie, Auf dem H\"ugel 69, 53121 Bonn, Germany
(l.loinard@crya.unam.mx).
\item Alejandro C. Raga: Instituto de Ciencias Nucleares, Universidad Nacional Aut\'onoma de M\'exico, Apdo. 
Postal 70-543, CP. 04510, D. F., M\'exico
(raga@nucleares.unam.mx).
\item Jorge Cant\'o: Instituto de Astronom\'\i a, Universidad Nacional Aut\'onoma de M\'exico, Apdo.
Postal 70-264, CP. 04510, D. F., M\'exico
\item Angels Riera: Departament de F\'\i sica i Enginyeria Nuclear, EUETIB, Universitat Polit\`ecnica de Catalunya, 
Compte d'Urgell 187, 08036 Barcelona, Spain
}

\shortauthor{Rodr\'\i guez et al.}
\shorttitle{Radio and Optical Observations of DG Tau B}

\SetVolume{50} \SetFirstPage{001} \SetYear{2012}

\resumen{
DG Tau B es un objeto estelar joven de Clase I que produce el chorro bipolar
asim\'etrico HH 159. A longitudes de onda visibles est\'a oscurecido por material
circunestelar \'opticamente grueso. Usando observaciones de VLA y el JVLA determinamos
por primera vez los movimientos propios de esta fuente y los encontramos consistentes,
dentro del error,
con los de la estrella cercana DG Tau. Tambi\'en discutimos un evento de eyecci\'on que es
evidente en los datos del VLA de 1994. Al igual que los flujos \'optico y molecular,
esta eyecci\'on observada en el radiocontinuo es marcadamente asim\'etrica 
y se detect\'o s\'olo hacia el noroeste de la estrella. Proponemos que este nudo,
que ya no es detectable en el radio, podr\'\i a ser observado en futuras im\'agenes \'opticas
de DG Tau B. Las posiciones de la fuente VLA y de un objeto infrarrojo cercano
no son coincidentes espacialmente
y sugerimos que es la fuente VLA la que traza al objeto excitador, mientras que
la fuente infrarroja podr\'\i a ser un l\'obulo de reflexi\'on.
}
 
\abstract{
DG Tau B is a Class I young stellar source that drives the asymmetric HH 159 bipolar jet. 
At optical wavelengths it is obscured by circumstellar optically-thick material.
Using VLA and JVLA observations,
we determine for the first time the proper motions of this source and find
them to be consistent, within error, with those of the nearby young star DG Tau.
We also discuss an ejection event that is evident in the 1994 VLA data. As the optical
and molecular outflows, this ejection traced in the radio continuum is
markedly asymmetric and was detected only to the NW of the star.
We propose that this knot, no longer detectable in the radio, could
be observed in future optical images of DG Tau B.
The positions of the VLA source and of a nearby infrared object 
are not coincident and we suggest that the VLA source traces
the exciting object, while the infrared source could be 
a reflection lobe.
}
      
\keywords{ISM: JETS AND OUTFLOWS --- STARS: INDIVIDUAL (DG TAU B) --- RADIO CONTINUUM: STARS}

\begin{document}

\maketitle

\section{Introduction}

DG Tau B is a young stellar object that has been classified as a Class I 
source based on its spectral energy distribution (Watson et al. 2004; Luhman et al. 2010).
Its luminosity is estimated to be 0.88 $L_\odot$ (Jones \& Cohen 1986).
DG Tau B is located in the sky approximately in between the
L1495 region and the star HP Tau. There are accurate distance
determinations from very long baseline interferometry geometric
parallax to both the L1495 region (131.5 pc; Torres et al.
2007; 2012) and HP Tau (161 pc; Torres 2009). Here we adopt
for DG Tau B a distance of 150 pc, intermediate to those of L1495
and HP Tau.
At optical wavelengths it is obscured by circumstellar optically thick material detected 
in absorption through broadband imaging with the Hubble Space telescope (Stapelfeldt et al. 1997). 
It drives the very asymmetric bipolar jet HH 159, first detected by Mundt \&  Fried (1983).
Millimeter $^{13}$CO observations (Padgett et al. 1999) indicate that the
obscuring material is
part of a circumstellar disk of elongated morphology and aligned perpendicular to 
the axis of HH 159.

While several jets from young stars exhibit considerable symmetry (e.g. HH 212; Correia et
al. 2009), the HH 159 jet has always
been remarkable for its asymmetry (Podio et al. 2011). The red optical lobe consists of a chain of 
bright knots extending to  $\sim55{''}$ to the NW of the source, while the blue 
optical lobe is fainter 
and less collimated and is detected only up to $\sim10{''}$ to the SE
of the source (Mundt et al. 1991; Eisl\"offel \& Mundt 1998). McGroarty 
\& Ray (2004) proposed that the Herbig-Haro
objects 836 and 837, located several arcmin to the SE of DG Tau B, could be
tracing ejecta from this star that took place $\sim10^4$ years ago. A large molecular outflow 
spatially coincident with the redshifted optical jet has been detected in the CO 
lines by Mitchell et al. (1994, 1997). These authors find that the CO outflow is
similar to the optical outflow: the CO redshifted emission extends at least 6000 AU
($40{''}$ at the distance of 150 pc)
to the NW of the star while the blueshifted CO emission is confined to a compact region, 
which is less than 500 AU ($\sim3{''}$) in radius.
Mitchell et al. (1997) suggest that a molecular clump or core surrounds DG Tau B.

DG Tau B was first detected in the radio continuum by Rodr\'\i guez et al. (1995),
who found a sub-arcsecond elongated source (with deconvolved angular
dimensions of $0\rlap.{''}38 \pm 0\rlap.{''}02 \times 
0\rlap.{''}22 \pm 0\rlap.{''}02$), with its major axis aligned along a position 
angle of $298^\circ \pm 5^\circ$, coincident within a few degrees with the 
position angle of the axis of the optical and molecular outflows ($\sim294^\circ$).
These authors proposed that the radio emission was tracing the base of the collimated jet
responsible for the optical and molecular outflows. Using VLA observations
and data from the literature (Mitchell et al. 1997; Looney et al. 2000), the
study by Rodmann et al. (2006)  
finds a spectral index of $\alpha = 0.3$ ($S_\nu \propto \nu^{\alpha}$) in the 3.6 cm to 1.3 cm
range, and of $\alpha = 2.9$ in the 7 mm to 2.6 mm range. The spectral index in the centimeter
range is consistent with a thermal (free-free) jet, while that in the millimeter
range indicates optically-thin emission from dust.

In this paper we use archive data from the Very
Large Array (VLA) and the Jansky Very Large Array (JVLA)
of the NRAO\footnote{The National Radio
Astronomy Observatory is operated by Associated Universities
Inc. under cooperative agreement with the National Science Foundation.}
to study the flux density, morphology, and proper motions of the radio
source associated with DG Tau B in the period since its
detection in 1994 VLA data (Rodr\'\i guez et al. 1995).
We also use optical images to search for newly ejected knots
and to discuss the identification of
the exciting source of the system.

\section{Radio Observations}

The archive epochs used are listed in Table 1.
All epochs, except the last one, were obtained with the VLA at 3.6 cm using
a 100 MHz effective bandwidth, and were
edited and calibrated using the software package Astronomical Image
Processing System (AIPS) of NRAO.
The last epoch is the average of three observations made
at 4.0 cm with the JVLA (Jansky Very Large Array)
in 2011 February 27, March 20, and May 5. These observations were made
with an effective bandwidth of 1 GHz and are
part of the Gould's Belt Distance Survey (Loinard et al. 2011).
They were edited and calibrated with the software package
Common Astronomy Software Applications (CASA).
The full set of observations covers about 17 years.

\begin{table}[htbp]
\small
  \setlength{\tabnotewidth}{1.0\columnwidth} 
    \tablecols{4} 
      \caption{Flux densities of DG Tau B at 3.6 cm}
	\begin{center}
	    \begin{tabular}{lccc}\hline\hline
	       &  Flux Density   & VLA/JVLA & VLA/JVLA \\
	       Epoch  & (mJy) & Configuration & Project \\
	       \hline
        1994 Apr 16 & 0.30$\pm$0.05$^a$ & A & AR277 \\ 
	1994 Jul 08 & 0.29$\pm$0.05 & B & AW386 \\
	1996 Dec 24 & 0.32$\pm$0.04 & A & AR277  \\
	1997 Dec 15 & 0.44$\pm$0.08 & D & AW472  \\
	1999 Nov 18 & 0.32$\pm$0.05 & B & AW522  \\
	2003 Mar 21 & 0.36$\pm$0.04 & D & AM756  \\
	2009 Aug 13 & 0.34$\pm$0.02 & C & AR694  \\
	2011 Apr 01$^b$ & 0.29$\pm$0.01$^c$ & B & BL175 \\
	       \hline\hline
\tabnotetext{a}{This flux density comes from the component associated
with the star. The ejected knot to the NW has a flux density of 0.26$\pm$0.05 mJy.}
\tabnotetext{b}{Mean epoch of the three epochs observed with the JVLA: 2011 February 27, 2011 March 20,
and 2011 May 05.}
\tabnotetext{c}{Flux density at 4.0 cm from the concatenated data of the three epochs.}
   \label{tab1}
      \end{tabular}
 \end{center}
\end{table}

\subsection{Flux density as a function of time}

In Table 1 we show the flux densities of DG Tau B as a function of time.
The flux density appears to be approximately constant, with a weighted mean
and weighted standard deviation given by
0.305$\pm$0.026 mJy. To test statistically the possibility of
time variation, we did a $\chi^2$ fitting to the 8 data points
assuming they are modeled by a constant with a flux density of 0.305 mJy.
We obtain a value of $\chi^2$ = 10.4, which is somewhat high,
but still consistent with the
range expected for a good fit, which is given by
$\nu \pm \sqrt {2 \nu}$ 
(Lightman et al. 1992), with $\nu$ being the number of degrees of freedom.
In our case $\nu$ = 7 and the range of $\chi^2$ for a good fit is
7.0$\pm$3.7.
We then conclude
that DG Tau B does not show significant flux density variations over the
time interval covered by the observations.

\subsection{Proper motions}

There are several limitations to undertake the astrometry of a radio
source from VLA and JVLA archive observations. The observations have to be made
with good signal to noise ratios and the best angular resolution possible.
In addition, the same phase calibrator should have been used in the observations.
With these limitations, we found only five epochs of data
(1994 April 16, 1994 July 8, 1994 December 24, 2009 August 13, and 2011 April 1), four with the
VLA and the last one with the JVLA that could be used. All observations
used J0403+260 as phase calibrator. 
The position of DG Tau B as a function of
time is shown in Figure 1.
For the first epoch we assumed that the position of the star
was that of the SE component (see Figure 2 and discussion below).
The proper motions derived from these
positions are

\begin{eqnarray}
\mu_\alpha\cos \delta  & = &  +3.8 \pm 1.9 \mbox{~mas~yr$^{-1}$}\nonumber\\
\mu_\delta  & = &  -20.6 \pm 3.3 \mbox{~mas~yr$^{-1}$.} \nonumber
\end{eqnarray}

The correlation coefficients for these
fits are 0.67 for the right ascension and -0.95 for the
declination. These proper motions are consistent within 2-$\sigma$ with those reported
by Rodr\'\i guez et al. (2012) for the nearby young star DG Tau (located $\sim 54''$ to the
NE of DG Tau B). This result confirms that, as expected, DG Tau and DG Tau B share
very similar proper motions since they are part of the same star-forming region.
There are no proper motions for DG Tau B from other wavelengths since it is a
heavily obscured object.

\begin{figure}
   \centering
 \includegraphics[scale=0.40]{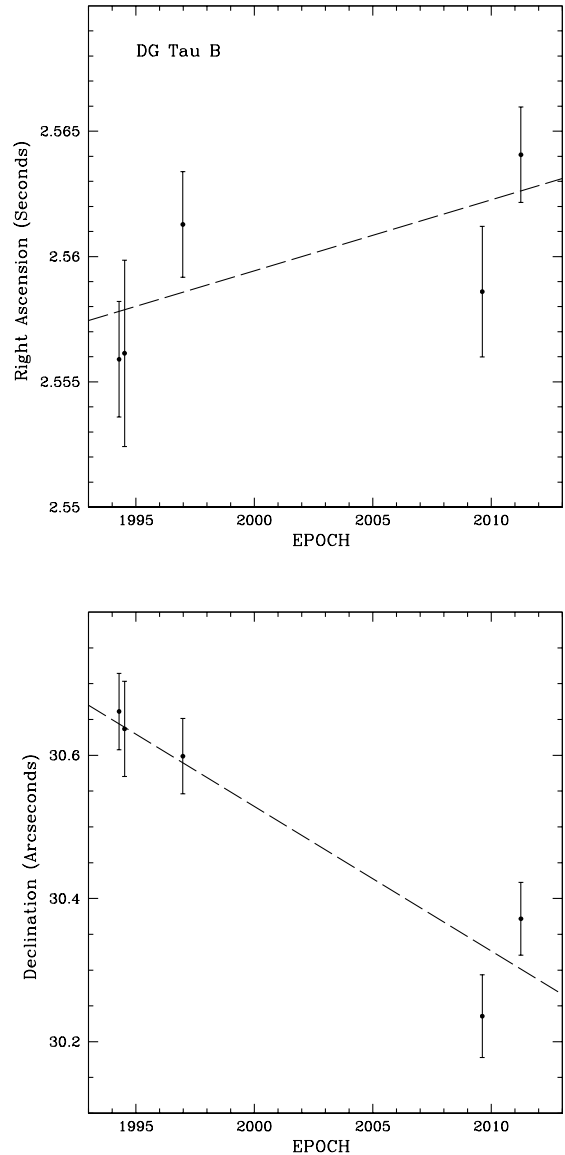}
 \caption{\small Position of the radio emission of DG Tau B
as a function of time from VLA and JVLA data. The right ascension (top) is 
given with respect to $04^h~27^m$
and the declination (bottom) with respect to $+26^\circ~05'$.
The dashed lines are least-squares fits to the data that
give the proper motions discussed in the text.}
\label{propermotions}
\end{figure}

\subsection{Morphology}

The image of Rodr\'\i guez et al. (1995) clearly shows an elongated source.
However, the stellar position of DG Tau B was poorly known then and it was
not possible to accurately position the star with respect to the radio
emission. Our reanalysis of the data confirms the elongated structure
and suggests that we are actually observing a double source (see Figure 2).
It is possible to fit this structure with two sources separated by
$0\rlap.{''}21 \pm 0\rlap.{''}04$ and with 3.6 cm flux densities of
0.30$\pm$0.05 mJy and 0.22$\pm$0.05 mJy for the SE and NW components, respectively. 
We assumed that the SE component is the one associated with the star.
We believe that this assumption is correct since this position is consistent
with the rest of the astrometry, with the other four epochs not showing evidence
for an additional component.
We conclude that the NW component seen in 1994 (see top panel of Figure 2)
is probably an ejecta that took place somewhere in the past.
The 1996 December 24 image (see bottom panel of Figure 2), of very similar angular resolution
and sensitivity as the 1994 April 16 image, does not show the ejecta at a level
$\sim$4 times lower than that present in 1994.

\begin{figure}
\centering
\includegraphics[scale=0.45, angle=0]{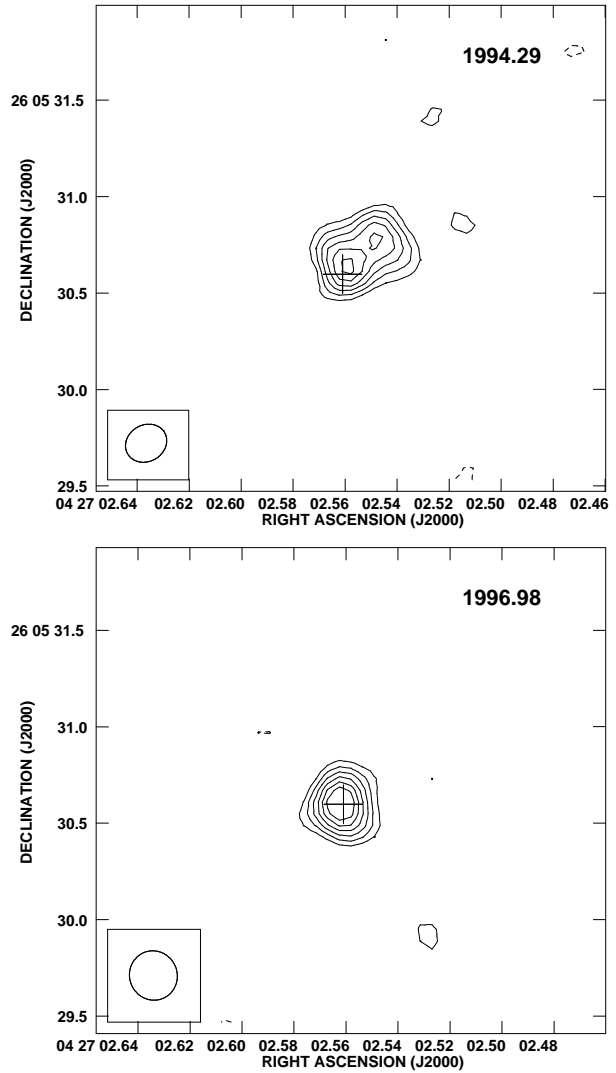}
 \caption{VLA contour images of the 3.6-cm continuum emission toward
DG Tau B. (Top) Image for 1994 April 16. Contours are -3, 3, 4, 5, 6, 7, and 8
times 0.015 mJy beam$^{-1}$, the rms noise of the image. 
The synthesized beam, shown in the bottom left corner,
has half power full width dimensions of
$0\rlap.{''}22 \times 0\rlap.{''}19$, 
with the major axis at a position angle of $-60^\circ$. 
(Bottom) Image for 1996 December 24. Contours are -3, 3, 4, 5, 6, 7, and 8
times 0.015 mJy beam$^{-1}$, the rms noise of the image.
The synthesized beam, shown in the bottom left corner,
has half power full width dimensions of
$0\rlap.{''}26 \times 0\rlap.{''}25$,
with the major axis at a position angle of $+21^\circ$.
The cross marks the radio position of DG Tau B in this second epoch,
$\alpha(2000) = 04^h~ 27^m~ 02\rlap.^{s}561; \delta(2000) = +26^\circ~ 05'~ 30\rlap.{''}60$.
}
 \label{fig2}
\end{figure}

The radio ejecta from young stellar objects are known to decrease in flux density with time
and eventually become undetectable in timescales of years (e.g. Mart\'\i\ et al. 1998).
We then find the non detection of the knot in the 1996 data consistent with
what is known of radio knots from young stellar objects.
However, we now address the following question: is the radio knot observed in 1994 evident
in more recent optical images of the HH 159 jet? The optical observations are much more
sensitive for the detection of knots than the radio ones. In the nearby young star DG Tau
Rodr\'\i guez et al. (2012) find that the radio knots are, within positional
error, coincident with optical knots.

>From the observations made between 1986 and 1990, Eisl\"offel \& Mundt (1998) 
determine an average
proper motion of 0.12$\pm$0.03 arcsec yr$^{-1}$ (87$\pm$22 km s$^{-1}$) away from the star
for the five inner optical knots of the NW lobe (knots A1, A3, A2, B1, and B2, in order
of increasing distance from the star). If we assume that the radio knot had this proper motion,
we conclude that it was ejected about 1.8 years before the 1994.29 observations, that is around
1992.5.  Are there more recent optical observations of DG Tau B that reveal the appearance
of a new knot? The most recent optical observations of the optical knots of DG Tau B
reported in the literature
were taken on 1998 February 23 and 24 with the Keck I telescope (Podio et al. 2011).
In these data the closest knot is A1, at $\sim3{''}$ from the star. This is the same
knot reported by Eisl\"offel \& Mundt (1998) as the closest to the star in data 
taken a decade before.
>From the above assumptions, the radio knot was expected to be at $\sim0\rlap.{''}7$
from the star in early 1998. It is then not surprising that it was not detected 
in the optical observations reported by Podio et al. (2011) because
there is heavy obscuration in the NW optical lobe within $\sim1\rlap.{''}2$ from the star
(Stapelfeldt et al. 1997).
However, at present (2012.0), the knot should be located at $2\rlap.{''}3$ from the star and is possibly
detectable in new optical images.

\begin{figure}
   \centering
 \includegraphics[scale=0.45]{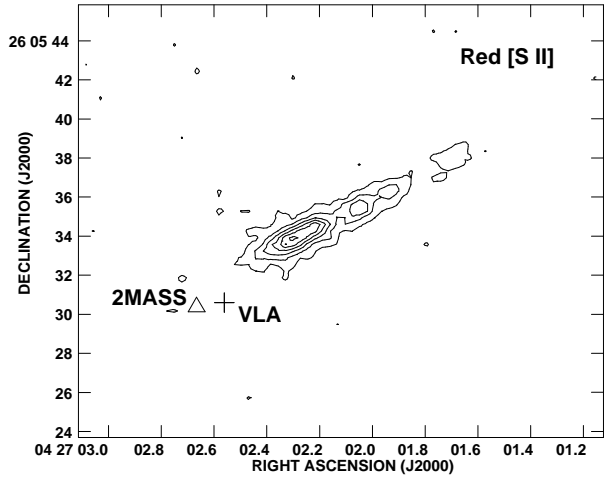}
\caption{\small Red [S II] image of the DG Tau B outflow.
Contours are 15, 16, 17, 18, 19, and 20 times 101, in arbitrary
units. 
The cross indicates the position of the VLA source and the triangle
the position of 2MASS J04270266+2605304. The astrometry of the [S II] image
has an accuracy of $\sim1{''}$.}
\label{[SII]}
\end{figure}

\section{Optical Observations}

We have obtained an image of this object in the night of
February 23, 2010. The narrow band image of DG Tau was obtained
at the 2.6m Nordic  Optical Telescope (NOT) of the Roque de Los Muchachos 
Observatory (La Palma, Spain)
using the Service Time mode facility. The image was obtained with 
the Andalucia Faint Object Spectrograph and Camera (ALFOSC) 
in imaging mode. The detector was an E2V 2Kx2K
CCD with a pixel size of 13.5$\mu$m, providing a plate scale 
of 0.19 arcsec pixel$^{-1}$. A [S II] filter (central wavelength 
$\lambda$ = 6725 \AA, FWHM = 60 \AA) was used to obtain an image 
of DG Tau in the [S II] 6716, 6731 \AA~ emission lines. 

Two exposures of 900 seconds each were combined to obtain the final 
image. The angular resolution during the
observations, as derived from the FWHM of stars in the field of 
view, was of 0.9 - 1.0 arcsec. 
The images were processed with the standard tasks of the IRAF reduction 
package. 

\subsection{Astrometry}

The [S~II] image is shown in Fig. 3. To locate the radio source in the optical image 
and discuss the relation between radio and optical sources,
we did the astrometry of the image using the task XTRAN in AIPS. Since there were
only four stars in the field, the astrometry is of modest quality, with
an absolute positional error of order $\sim 1{''}$.
In this image we have put the positions of the VLA source
($\alpha(2000) = 04^h~ 27^m~ 02\rlap.^{s}561; \delta(2000) = +26^\circ~ 05'~ 30\rlap.{''}60$)
and of the infrared source 2MASS J04270266+2605304
($\alpha(2000) = 04^h~ 27^m~ 02\rlap.^{s}666; \delta(2000) = +26^\circ~ 05'~ 30\rlap.{''}38$),
this last position taken from Skrutskie et al.(2006).
This infrared source is usually considered to be the exciting source
of the DG Tau B outflow. 
There is no [S II] emission within $2\rlap.{''}3$ of the VLA or the 2MASS source.
Under the hypothesis that the radio knot moves with the average
proper motion of the five inner optical knots,
we can conclude that either our hypothesis that the radio knot should be eventually
detected in the optical is wrong or that extinction is still very strong
at these distances from the star and the knot will become visible only in the future. 
However, the radio knot could be moving at faster or slower proper motions.
If it is moving much slower, it could still be hidden in the highly
obscured region. If it was moving much faster
it could, however, have caught up with the first old optical knot A1.
This could lead to changes of the morphology and brightness of this knot in the
new image, compared to the old images. However, the quality of the data does
not allow a detailed comparison between images.

\subsection{Which object is the exciting source of the DG Tau B outflow?}

Interestingly, the positions of the VLA source and 2MASS J04270266+2605304 do not coincide, with
the latter source being $\sim 1{''}$ to the east of the former.
To investigate this issue further, we superposed the positions of these sources
on the red image of the Second Digitized Sky Survey (Fig. 4).
This optical image shows emission associated with the NW jet but also emission
to the SE of the compact sources. While the VLA source seems to fall
in the darkest part of the structure, 2MASS J04270266+2605304 appears
to coincide with the SE structure. Tentatively, we propose that
the VLA source traces the true exciting source, while 2MASS J04270266+2605304
could be an infrared reflection nebula associated with the SE lobe of the structure.
This result is consistent with the conclusion of Mitchell et al. (1997),
who found that the peak of the 2.6 mm dust continuum emission
was displaced by $\sim 1\rlap.{''}2$ to the NW of the ``stellar'' position of DG Tau B
(Mundt et al. 1987).

\begin{figure}
   \centering
  \includegraphics[scale=0.45]{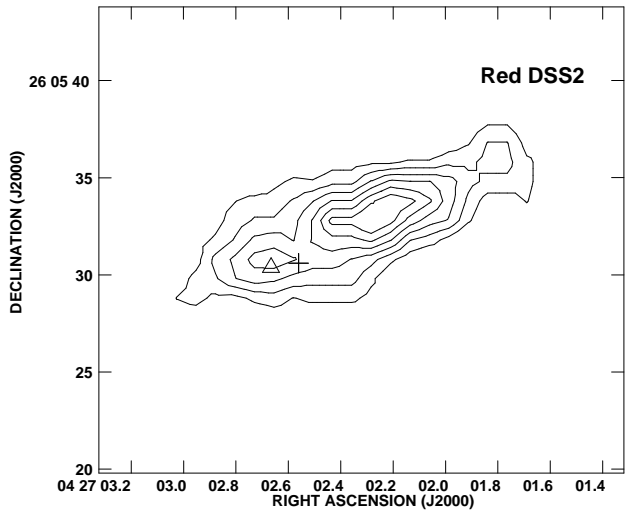}
  \caption{\small Red image of the Second Digitized Sky Survey
  for the the DG Tau B region.
Contours are 5.5, 6.0, 6.5, 7.0, 7.5, and 8.0 times 1010, in arbitrary
units. 
The cross indicates the position of the VLA source and the triangle
the position of 2MASS J04270266+2605304.}
\label{RedDSS2}
\end{figure}

Also favoring the VLA object as the exciting source of the outflow is the
fact that the accurate position of the dust millimeter emission reported by
Guilloteau et al. (2011), that traces a disk around the star, coincides
with the VLA position within $\sim 0\rlap.{''}1$.

\section{Conclusions}

We have analyzed VLA and JVLA archive data to determine the proper motions of DG Tau B. These proper
motions are consistent within error
with those determined for the nearby young star DG Tau (Rodr\'\i guez et al. 2012).
We show that the elongated radio structure observed by Rodr\'\i guez et al. (1995) 
corresponds to emission associated with the star plus a knot ejected in mid-1992. 
We propose that if this radio knot, no longer detectable in the radio, is producing
detectable optical emission it could be observed as a ``new'' optical knot in 
future images.

There are two sources at the center of the outflow, one is the VLA source
studied here and the other is the infrared source 2MASS J04270266+2605304.
We propose that the true exciting source is the VLA object, while
the infrared source is a reflection nebula. 


\acknowledgments
L.F.R., S.A.D., L.A.Z., A.C.R. and J.C. are thankful for the support
of DGAPA, UNAM, and of CONACyT (M\'exico).
A.R. is partially supported by Spanish MCI grants
AYA2008-06189-C03 and AYA2011-30228-C03, and FEDER funds.
L.L. is indebted to the Alexander von Humboldt Stiftung and the Guggenheim Memorial 
Foundation  for financial support.
This research has made use of the SIMBAD database, 
operated at CDS, Strasbourg, France.
The Digitized Sky Survey image used in this
paper was produced at the Space Telescope Science Institute 
under U.S. Government grant NAG W-2166. 
The images of these surveys are based on photographic data obtained using the Oschin Schmidt 
Telescope on Palomar Mountain and the UK Schmidt Telescope. The plates were processed into 
the present compressed digital form with the permission of these institutions.
This paper is partially based on observations made
with the Nordic Optical Telescope, operated on the island of La Palma
jointly by Denmark, Finland, Iceland, Norway, and Sweden, in the Spanish
Observatorio del Roque de los Muchachos of the Instituto de Astrof\'\i sica de
Canarias. 

\vskip0.5cm



\begin{thebibliography}

\bibitem{2009A&A...505..673C} Correia, S., Zinnecker, H., Ridgway, S.~T., \& 
McCaughrean, M.~J.\ 2009, \aap, 505, 673 

\bibitem{1998AJ....115.1554E} Eisl{\"o}ffel, J., \& Mundt, R.\ 1998, \aj, 115, 1554 

\bibitem{2011A&A...529A.105G} Guilloteau, S., Dutrey, A., Pi{\'e}tu, V., \& Boehler, 
Y.\ 2011, \aap, 529, A105 

\bibitem{1986ApJ...311L..23J} Jones, B.~F., \& Cohen, M.\ 1986, \apjl, 311, L23 


\bibitem{2011RMxAC..40..205L} Loinard, L., 
Mioduszewski, A.~J., Torres, R.~M., et al.\ 2011, Revista Mexicana de 
Astronomia y Astrofisica Conference Series, 40, 205 

\bibitem{2000ApJ...529..477L} Looney, L.~W., Mundy, 
L.~G., \& Welch, W.~J.\ 2000, \apj, 529, 477 

\bibitem{2010ApJS..186..111L} Luhman, K.~L., Allen, 
P.~R., Espaillat, C., Hartmann, L., \& Calvet, N.\ 2010, \apjs, 186, 111 

\bibitem{1998ApJ...502..337M} Marti, J., Rodriguez, 
L.~F., \& Reipurth, B.\ 1998, \apj, 502, 337 

\bibitem{2004A&A...420..975M} McGroarty, F., \& Ray, T.~P.\ 2004, \aap, 420, 975 

\bibitem{1994ApJ...436L.177M} Mitchell, G.~F., 
Hasegawa, T.~I., Dent, W.~R.~F., \& Matthews, H.~E.\ 1994, \apjl, 436, L177 

\bibitem{1997ApJ...483L.127M} Mitchell, G.~F., 
Sargent, A.~I., \& Mannings, V.\ 1997, \apjl, 483, L127 

\bibitem{1983ApJ...274L..83M} Mundt, R., \& Fried, J.~W.\ 1983, \apjl, 274, L83 

\bibitem{1987ApJ...319..275M} Mundt, R., Brugel, E.~W., 
\& Buehrke, T.\ 1987, \apj, 319, 275 

\bibitem{1991A&A...252..740M} Mundt, R., Ray, T.~P., \& Raga, A.~C.\ 1991, \aap, 252, 740 

\bibitem{1999AJ....117.1490P} Padgett, D.~L., 
Brandner, W., Stapelfeldt, K.~R., et al.\ 1999, \aj, 117, 1490 

\bibitem{2011A&A...527A..13P} Podio, L., Eisl{\"o}ffel, J., Melnikov, S., 
Hodapp, K.~W., \& Bacciotti, F.\ 2011, \aap, 527, A13 

\bibitem{1992nrfa.book.....P} Press, W.~H., Teukolsky, 
S.~A., Vetterling, W.~T., 
\& Flannery, B.~P.\ Numerical recipes in FORTRAN. The art of scientific computing.
1992, Cambridge: University Press, 2nd ed.

\bibitem{2006A&A...446..211R} Rodmann, J., Henning, T., Chandler, C.~J., 
Mundy, L.~G., \& Wilner, D.~J.\ 2006, \aap, 446, 211 

\bibitem{1995ApJ...454L.149R} Rodr\'\i guez, L.~F., 
Anglada, G., \& Raga, A.\ 1995, \apjl, 454, L149 

\bibitem{2012A&A...537A.123R} Rodr{\'{\i}}guez, L.~F., Gonz{\'a}lez, R.~F., Raga, A.~C., 
et al.\ 2012, \aap, 537, A123 

\bibitem{2006AJ....131.1163S} Skrutskie, M.~F., 
Cutri, R.~M., Stiening, R., et al.\ 2006, \aj, 131, 1163 

\bibitem{1997IAUS..182..355S} Stapelfeldt, K., 
Burrows, C.~J., Krist, J.~E., 
\& WFPC2 Science Team 1997, Herbig-Haro Flows and the Birth of Stars, 182, 355

\bibitem{torres}
Torres, R.~M., Loinard,
L., Mioduszewski, A.~J., \& Rodr{\'{\i}}guez, L.~F.\ 2007, ApJ, 671, 1813

\bibitem{2009ApJ...698..242T} Torres, R.~M., Loinard, 
L., Mioduszewski, A.~J., \& Rodr{\'{\i}}guez, L.~F.\ 2009, \apj, 698, 242 

\bibitem{2012ApJ...747...18T} Torres, R.~M., Loinard, 
L., Mioduszewski, A.~J., et al.\ 2012, \apj, 747, 18 

\bibitem{2004ApJS..154..391W} Watson, D.~M., Kemper, 
F., Calvet, N., et al.\ 2004, \apjs, 154, 391 




\end{thebibliography}
\end{document}